\documentclass[showpacs,twocolumn,typeset,superscriptaddress]{revtex4}

\usepackage{etoolbox} 
\usepackage{lipsum} 
\usepackage[capitalize]{cleveref}

\usepackage{amsfonts}
\usepackage{amsmath}
\usepackage{amssymb}
\usepackage{graphicx}
\usepackage{dcolumn}
\usepackage{bm}

\begin{document}

\title{A noisy channel less entangled than a pure can improve teleportation}

\author{Luis Roa}\email{lroa@udec.cl}
\affiliation{Departamento de F\'{\i}sica, Universidad de Concepci\'{o}n, Casilla 160-C, Concepci\'{o}n, Chile.}
\author{M. Loreto Ladr\'on de Guevara}
\affiliation{Departamento de F\'{\i}sica, Facultad de Ciencias, Universidad Cat\'olica del Norte, Casilla 1280, Antofagasta, Chile.}
\author{Matias Soto-Moscoso}
\affiliation{Departamento de F\'{\i}sica, Universidad de Concepci\'{o}n, Casilla 160-C, Concepci\'{o}n, Chile.}
\affiliation{Departamento de F\'{\i}sica, Facultad de Ciencias, Universidad del B\'{\i}o-B\'{\i}o, Avenida Collao 1202, Casilla 15-C, Concepci\'on, Chile.}
\author{Pamela Catal\'an}
\affiliation{Departamento de F\'{\i}sica, Universidad de Concepci\'{o}n, Casilla 160-C, Concepci\'{o}n, Chile.}
\affiliation{Departamento de F\'{\i}sica, Facultad de Ciencias, Universidad del B\'{\i}o-B\'{\i}o, Avenida Collao 1202, Casilla 15-C, Concepci\'on, Chile.}
\affiliation{Facultad de Ingenier\'{\i}a y Tecnolog\'{\i}a, Universidad San Sebasti\'an, Campus Las Tres Pascualas, Lientur 1457, Concepci\'on, Chile.}
\date{\today}

\begin{abstract}
In our work we consider the following problem in the context of teleportation:
an unknown pure state have to be teleported and there are two laboratories which can perform the task.
One laboratory uses a pure non maximally entangled channel but has a capability of performing the joint measurement on bases with a constrained degree of entanglement;
the other lab makes use of a mixed $X$-state channel but can perform a joint measurement on bases with higher entanglement degrees.
We compare the average teleportation fidelity achieved in both cases, finding that the
fidelity achieved with the $X$-state can surpass the obtained with a pure channel, even though the $X$-state is less entangled than the latter.
We find the conditions under which this effect occurs.
Our results evidence that the entanglement of the joint measurement plays a role as important as the entanglement of the channel in order to optimize the teleportation process.
We include an example showing that the average fidelity of teleportation obtained with a Werner state channel can be grater than that obtained with a Bell state channel.
\end{abstract}

\pacs{03.67.Bg, 03.65.Ud, 03.67.Ac} \maketitle

\section{Introduction}

Teleportation is one of those special processes in quantum information theory without classical counterpart.
The ideal teleportation process is able to transmit unknown quantum information from a quantum system to another similar without any direct interaction between them,
no matter how distant they are \cite{GBrassard,Bowmeester,Boschi,Caves}.
A key resource for teleporting successfully, i.e., with fidelity $1$, is a quantum channel consisting of a bipartite system prepared in a maximally entangled state \cite{GBrassard}.
Other important ingredients are the ability of performing a joint-measurement of maximally entangled states and the capacity of communicating the result by means of a classical channel.
With these resources the teleportation becomes deterministic.
However, more realistic teleportation descriptions must take into account the undesired mechanisms which introduce decoherence, dephasing, and dissipation on the quantum channel,
leading it from a maximally entangled pure state to a partially entangled mixed state \cite{Wilhelm}.
Thus, due to the environment, an initial Bell state can evolve into a stationary $X$-state, which can preserve partially some quantum features \cite{Man,Yu}.
Therefore, the latter family of these states is a good option for representing a real noisy quantum channel.
On the other hand, to perform a Bell joint-measurement is not possible in some cases \cite{Lutkenhaus}.
In this context, non-Bell channel and joint measuring-bases have been introduced,
finding the entanglement-matching effect \cite{Wan-Li} and the probabilistic successful teleportation with or without loss of information \cite{Banaszek}.

More general teleportation schemes considering these undesired aspects have been studied \cite{Wan-Li,Banaszek,Bandyopadhyay,Brassard,Popescu,Cavalcanti}.
In particular, it has been found that the average fidelity of the deterministic teleportation becomes larger than $2/3$ when the mixed state channel has non-classical features \cite{Popescu,Gisin,Verstraete,lroa}.
Previous to apply the teleportation scheme, the mixed state channel can be manipulated via local filtering operations with the purpose of obtaining a state leading to the highest possible average teleportation fidelity, which has a Bell-diagonal normal form \cite{Verstraete}.
The drawback of this filtering process is the fact that it is implemented with a success probability different from $1$, e.i., there is a non-zero probability of losing the quantum correlation of the channel.

In this work we show that, in the teleportation procedure, the entanglement of the joint measurement plays a role as important as the entanglement of the channel.
Specifically, it is crucial in the sense that the average fidelity of the teleportation outcomes reached with a noisy channel in an $X$-state can be improved
with respect to the obtained with a pure state channel, even though the entanglement of the noisy channel is smaller.
This is the main result of our work.
This scheme does not include any probabilistic filtering process onto the channels.
However, the considered $X$-state channels have in some especial cases the Bell-diagonal normal form.

The article is organized as follows.
In Sec. \ref{sec2} we obtain the average fidelity in a teleportation scheme which uses a quantum channel consisting of two qubits in a partially entangled pure state and a joint-measurement with a non-Bell basis.
In Sec. \ref{sec3} we address the average fidelity in the teleportation scheme with an $X$-state channel and a non-Bell measurement basis. Here we prove the main result of this article.
Finally, in the last section, we summarize our principal results.

\section{Teleportation with a pure state channel} \label{sec2}

We start by reviewing the teleportation scheme performed with a pure and partially entangled quantum channel and a non maximally entangled measurement basis \cite{Wan-Li}.
Then we include an analysis of the average fidelity of the outcomes as a function of the involved concurrences.

We consider two distant qubits, $A$ and $B$, sharing the quantum channel in the following pure state,
\begin{equation}
|\psi _{AB}\rangle =\alpha |0\rangle |1\rangle +\beta |1\rangle |0\rangle ,
\label{purechannel}
\end{equation}%
where $\{|0\rangle ,|1\rangle\}$ are the eigenstates of the $\sigma_{z}$ Pauli operator.
The order of the subindexes is in accordance with the order of the subsystems in the tensor product terms.
Without loss of generality we assume that parameters $\alpha$ and $\beta$ are non-negative real numbers and $\alpha\leq\beta$.
This channel has an amount of entanglement characterized by the concurrence $C_{AB}=2\alpha\beta$ \cite{WW}.
To perform the teleportation scheme, instead of a Bell-measurement, we consider the joint-measurement performed onto the following orthonormal and partially entangled states,
\begin{subequations}
\begin{eqnarray}
|\phi _{+}\rangle  &=&x|0\rangle |0\rangle +y|1\rangle |1\rangle ,\\
|\phi _{-}\rangle  &=&y|0\rangle |0\rangle -x|1\rangle |1\rangle ,\\
|\psi _{+}\rangle  &=&x|0\rangle |1\rangle +y|1\rangle |0\rangle ,\\
|\psi _{-}\rangle  &=&y|0\rangle |1\rangle -x|1\rangle |0\rangle,
\end{eqnarray}%
\label{b}%
\end{subequations}%
where we have assumed that the parameters $x$ and $y$ are non-negative real numbers and $x\leq y$.
All the measurement basis states (\ref{b}) have the same amount of entanglement, given by the concurrence $C_{aA}=2xy$.
In this case, the teleportation of the unknown state $|\psi\rangle$, from qubit $a$ to qubit $B$, can be read from the following identity,
\begin{eqnarray}
|\psi\rangle_a |\psi_{AB}\rangle\!&=&\!\sqrt{p_{\phi _{+}}}|\phi _{+}\rangle\sigma_x|p_{\phi _{+}}\rangle\!-\!\sqrt{p_{\phi _{-}}}|\phi _{-}\rangle \sigma_{z}\sigma_x|p_{\phi _{-}}\rangle \nonumber \\
&&\!+\sqrt{p_{\psi _{+}}}|\psi _{+}\rangle|p_{\psi _{+}}\rangle\!-\!\sqrt{p_{\psi _{-}}}|\psi _{-}\rangle \sigma _{z}|p_{\psi_{-}}\rangle .  \label{id}
\end{eqnarray}%
Here we have defined the four normalized states of the qubit $B$,
\begin{subequations}
\label{p}
\begin{eqnarray}
|p_{\phi _{+}}\rangle  &=&\frac{x\alpha \langle 0|\psi \rangle |0\rangle
+y\beta \langle 1|\psi \rangle |1\rangle }{\sqrt{p_{\phi _{+}}}},  \label{p1}
\\
|p_{\phi _{-}}\rangle  &=&\frac{y\alpha \langle 0|\psi \rangle |0\rangle
+x\beta \langle 1|\psi \rangle |1\rangle }{\sqrt{p_{\phi _{-}}}},  \label{p2}
\\
|p_{\psi _{+}}\rangle  &=&\frac{x\beta \langle 0|\psi \rangle |0\rangle
+y\alpha \langle 1|\psi \rangle |1\rangle }{\sqrt{p_{\psi _{+}}}},
\label{p3} \\
|p_{\psi _{-}}\rangle  &=&\frac{y\beta \langle 0|\psi \rangle |0\rangle
+x\alpha \langle 1|\psi \rangle |1\rangle }{\sqrt{p_{\psi _{-}}}},
\label{p4}
\end{eqnarray}%
\end{subequations}%
where $p_{\phi _{\pm}}$ and $p_{\psi _{\pm}}$ are, respectively, the probabilities of projecting the qubits $aA$ onto the states $|\phi_\pm\rangle$ and $|\psi_\pm\rangle$.

From the right-hand side of identity (\ref{id}) we realize that there is a one to one and one-way correlation between the orthonormal states
$\{|\phi _{+}\rangle ,|\phi _{-}\rangle ,|\psi _{+}\rangle ,|\psi_{-}\rangle \}$ of the qubits $aA$ and the
linearly dependent states $\{\sigma_x|p_{\phi_{+}}\rangle ,\sigma _{z}\sigma_x|p_{\phi _{-}}\rangle ,|p_{\psi_{+}}\rangle ,\sigma _{z}|p_{\psi _{-}}\rangle \}$ of the qubit $B$.
Therefore, when projecting the qubits $aA$ onto one of the states (\ref{b}) the qubit $B$ is projected onto its correlated state.
The result of the measurement is sent, via classical communication, from laboratory $A$ to $B$,
information needed by the receiver to remove the suitable unitary, $\sigma _{x}$, $\sigma_z\sigma _{x}$, or $\sigma _{z}$, leaving thus the qubit $B$ in one of the states (\ref{p}).

Clearly, if $\alpha\neq \beta$ and $x\neq y$ none of the states (\ref{p}) is the state $|\psi\rangle$ to be teleported,
it is therefore natural to ask how similar are those to $|\psi\rangle$.
To quantify the distance between them we use the fidelity averaged over all of the outcomes (\ref{p}) and all of the possible states to be teleported, which is obtained from the functional \cite{Jozsa},
\begin{equation} \label{FG}
F= \frac{1}{2\pi^2}\int\sum_{k=\phi_{\pm},\psi _{\pm}}p_k\langle\psi|\rho_{p_k}|\psi\rangle d\psi,
\end{equation}
where, $d\psi$ is an infinitesimal volume element in the Hilbert space and, in this case, $\rho_{p_k}=|p_k\rangle\langle p_k|$.
Therefore, considering the outcomes (\ref{p}) and its probabilities, the average fidelity for teleportation becomes,
\begin{equation}
F_{p} = \frac{2}{3}+\frac{C_{aA}C_{AB}}{3}. \label{F1p}
\end{equation}%
From that expression we clearly obtain that $F_p$ is higher than $2/3$ iff both the concurrence of the channel and of the joint measurement are different from zero.
In other words, both entanglements are necessary and sufficient for observing the quantum nature of teleportation
in the average fidelity of the outcome states.
As expected, $F_p=1$ only when the measurement basis and the quantum channel are Bell states.

\section{Teleportation attempt with an X-State channel}    \label{sec3}

In this section we address the same scheme of teleportation, but instead of the pure channel (\ref{purechannel}),
we assume the qubits $AB$ share $\rho_{AB}$, an entangled $X$-state, which
in the bipartite logical basis $\{|0\rangle|0\rangle,|0\rangle|1\rangle,|1\rangle|0\rangle,|1\rangle|1\rangle\}$ has the matrix representation
\begin{equation}\label{Xstate}
\rho_{AB} \equiv \left(
\begin{array}{cccc}
\rho _{11} & 0 & 0 & \rho _{14} \\
0 & \rho _{22}  & \rho _{23}& 0 \\
0 & \rho _{32}  & \rho _{33}& 0 \\
\rho _{41} & 0 & 0 & \rho _{44}
\end{array}\right).
\end{equation}
We assume that the off-diagonal elements are real and non negative numbers, i.e., $\rho_{23}=\rho_{32}\geq 0$ and $\rho_{14}=\rho_{41}\geq 0$,
since their phases can be removed by local unitary transformations \cite{Gesa}.
The concurrence $\bar{C}_{AB}$ of the $X$-state (\ref{Xstate}) is well-known and given by \cite{Yu}
\begin{equation*}
\bar{C}_{AB}=2\max\{0,\rho_{23}-\sqrt{\rho_{11}\rho_{44}},\rho_{14}-\sqrt{\rho_{22}\rho_{33}}\}.
\end{equation*}
Taking into account that the $X$-state populates two orthogonal subspaces $\mathcal{H}_{0110}$ and $\mathcal{H}_{0011}$,
spanned by $\{|0\rangle|1\rangle,|1\rangle|0\rangle\}$ and $\{|0\rangle|0\rangle,|1\rangle|1\rangle\}$ respectively,
we can assume, without loss of generality, that $\mathcal{H}_{0110}$ contains the principal information, in the sense that it has the main term of the concurrence, i.e.,
$\rho_{23}-\sqrt{\rho_{11}\rho_{44}}>0\geq\rho_{14}-\sqrt{\rho_{22}\rho_{33}}$ \cite{Gesa}.
In consequence, $\rho_{AB}$ is entangled and its concurrence becomes
\begin{equation}
\bar{C}_{AB}=2(\rho_{23}-\sqrt{\rho_{11}\rho_{44}}).
\end{equation}
It is worth to realize that $\rho_{AB}$ takes the Bell-diagonal normal form when the populations inside each subspaces $\mathcal{H}_{0110}$ and $\mathcal{H}_{0011}$ are uniform, $\rho_{22}=\rho_{33}$ and $\rho_{11}=\rho_{44}$.
In this case $\rho_{AB}$ becomes,
\begin{eqnarray*}
\rho _{AB}&=&\left( \rho _{11}+\rho _{14}\right) |\tilde{\phi} _{+}\rangle \langle \tilde{\phi} _{+}|+\left( \rho _{11}-\rho _{14}\right) |\tilde{\phi} _{-}\rangle \langle \tilde{\phi} _{-}| \\
&&+\left( \rho _{22}+\rho _{23}\right) |\tilde{\psi} _{+}\rangle \langle \tilde{\phi} _{+}|+\left( \rho _{22}-\rho _{23}\right) |\tilde{\phi} _{-}\rangle \langle \tilde{\phi} _{-}|,
\end{eqnarray*}
where $|\tilde{\phi} _{\pm}\rangle=(|0\rangle|0\rangle\pm|1\rangle|1\rangle)/\sqrt{2}$ and $|\tilde{\psi} _{\pm}\rangle=(|0\rangle|1\rangle\pm|1\rangle|0\rangle)/\sqrt{2}$ are the Bell states.

To carry out the teleportation of the unknown state $|\psi\rangle$ from qubit $a$ to $B$, the sender performs the joint-measurement on the qubits $aA$ projecting them onto one of the states (\ref{b}).
Here we assume that each state of the measurement basis has concurrence $\bar{C}_{aA}$, which not necessarily has the same value than $C_{aA}$.
Once the sender has communicated the measurement result, the receiver can remove the suitable unitary to have one of the following states in its qubit $B$,
\begin{subequations}\label{rhosp}\begin{eqnarray}
\rho_{\phi_+} &=& \frac{\sigma_x\langle \phi_+|\left( |\psi \rangle \langle \psi |\otimes \rho_{AB}\right) |\phi_+\rangle\sigma_x}{p_{\phi_+}}, \\
\rho_{\phi_-} &=& \frac{\sigma_x\sigma_z\langle \phi_-|\left( |\psi \rangle \langle \psi |\otimes\rho_{AB}\right) |\phi_-\rangle\sigma_z\sigma_x}{p_{\phi_-}}, \\
\rho_{\psi_+} &=& \frac{\langle \psi_+|\left( |\psi \rangle \langle \psi |\otimes\rho_{AB}\right) |\psi_+\rangle}{p_{\psi_+}}, \\
\rho_{\psi_-} &=& \frac{\sigma_z\langle \psi_+|\left( |\psi \rangle \langle \psi |\otimes\rho_{AB}\right)|\psi_+\rangle\sigma_z}{p_{\psi_+}},
\end{eqnarray}
\end{subequations}
each one with the respective probability
\begin{subequations}\label{probofp}\begin{eqnarray}
p_{\phi_\pm} &=& Tr\langle\phi_{\pm}|\left(|\psi\rangle\langle\psi |\otimes\rho_{AB}\right)|\phi_{\pm}\rangle,   \\
p_{\psi_\pm} &=& Tr\langle\psi_{\pm}|\left(|\psi\rangle\langle\psi |\otimes\rho_{AB}\right)|\psi_{\pm}\rangle.
\end{eqnarray}\end{subequations}
By inserting the states (\ref{rhosp}) and their probabilities (\ref{probofp}) in the expression (\ref{FG}), we obtain the average fidelity $F_X$ for this teleportation process,
\begin{equation}\label{fidelityX_0}
F_X = \frac{2}{3}+\frac{\bar{C}_{aA}(\bar{C}_{AB}+2\sqrt{\rho_{11}\rho_{44}})-(\rho_{11}+\rho_{44})}{3}.
\end{equation}
We observe from Eq. (\ref{fidelityX_0}) that $\bar{C}_{AB}$  and $\bar{C}_{aA}$, basically, play the same roles in $F_X$,
since both concurrences contribute to increase the average fidelity.
The expression (\ref{fidelityX_0}) evidences the roles of the concurrences for generating the quantum effect and the populations belonging to the subspace $\mathcal{H}_{0011}$ for canceling it.
The off-diagonal element $\rho_{14}$ does not play any role in $F_X$.

At this point, there are two relevant questions which worth addressing.
What are the conditions fulfilled by the $X$-state channel and measurement basis under which the average fidelity displays quantum features?
Is it possible to make the average fidelity (\ref{fidelityX_0}) larger that its counterpart (\ref{F1p})
taking advantage of the fact that the second laboratory can realize joint measurements on bases with controllable entanglement amounts?
These issues are discussed in the following subsections.

\subsection{Threshold concurrences for quantum features}

We are interested in finding the conditions satisfied by $\rho_{AB}$ and $\bar{C}_{aA}$ under which the average fidelity $F_X$ displays quantum features, that is, $F_X>2/3$.
According to the expression (\ref{fidelityX_0}), this occurs when
\begin{equation}\label{threshA}
\bar{C}_{aA}(\bar{C}_{AB}+2\sqrt{\rho_{11}\rho_{44}})-(\rho_{11}+\rho_{44})>0.
\end{equation}
We assume first that $\bar{C}_{aA}\neq \bar{C}_{AB}$.
Under that condition, we solve (\ref{threshA}) for $\bar{C}_{AB}$ as a function of $\bar{C}_{aA}$, and vice versa.
This leads to the following inequalities,
\begin{equation}
\bar{C}_{AB} > \mathfrak{C}(1) \qquad\text{and}\qquad \bar{C}_{aA}>\mathcal{C}(\bar{C}_{AB}),   \label{Eq14}
\end{equation}
or
\begin{equation}
\bar{C}_{AB} > \mathfrak{C}(\bar{C}_{aA}) \qquad\text{and}\qquad \bar{C}_{aA} > \mathcal{C}(1),   \label{Eq15}
\end{equation}
where we have defined the functions
\begin{subequations}   \label{Eq16}
\begin{eqnarray}
\mathfrak{C}(c) &=& \frac{(\sqrt{\rho_{11}}-\sqrt{\rho_{44}})^2+2(1-c)\sqrt{\rho_{11}\rho_{44}}}{c},   \label{Eq16a}\\
\mathcal{C}(c)&=&\frac{\rho_{11}+\rho_{44}}{c+2 \sqrt{\rho_{11}\rho_{44}}}.  \label{Eq14c}
\end{eqnarray}
\end{subequations}
The inequalities (\ref{Eq14}) and (\ref{Eq15}) show that for $F_X$ to display quantum features, the channel and measurement basis concurrences,
$\bar{C}_{AB}$ and $\bar{C}_{aA}$,
must be greater than the threshold values,
distinct to what occurs when the channel is pure, where it is only required that $C_{AB}, C_{aA} > 0$.
In addition, the fact that $\mathfrak{C}(c)$ and $\mathcal{C}(c)$ are decreasing functions of $c$ implies that $\mathfrak{C}(1)<\mathfrak{C}(\bar{C}_{aA})$ and $\mathcal{C}(\bar{C}_{AB})>\mathcal{C}(1)$;
in consequence, the threshold entanglement of the channel in Eq. (\ref{Eq14}) is smaller than the one of Eq.(\ref{Eq15}),
whereas such relation is inverse for threshold values of the entanglement of the joint-measurement.
It is interesting to note that there are some special $X$-states for which one or all threshold conditions are released;
namely, the threshold value $\mathfrak{C}(1)$ vanishes when $\rho_{11}=\rho_{44}$, and all
the threshold values become zero when the subspace $\mathcal{H}_{0011}$ is not populated, i.e., when  $\rho_{11}=\rho_{44}=0$.

In the particular case where the concurrences of the channel and the joint-measurement are equal, the inequality (\ref{threshA}) is fulfilled only for
\begin{equation*}
\bar{C}_{AB}=\bar{C}_{aA}>\sqrt{\rho_{11}+\rho_{44}+\rho_{11}\rho_{44}}-\sqrt{\rho_{11}\rho_{44}}.
\end{equation*}
We notice that this threshold value vanishes only in the special case $\rho_{11}=\rho_{44}=0$.

\subsection{Threshold values for improving the average fidelity}

Let us now analyze the conditions fulfilled by $\rho_{AB}$ and $\bar{C}_{aA}$ in order that the average fidelity $F_X$ is larger than that obtained with the pure channel, $F_p$.
That occurs when the following inequality is satisfied,
\begin{equation}\label{threshX}
C_{aA}C_{AB}<\bar{C}_{aA}(\bar{C}_{AB}+2\sqrt{\rho _{11}\rho _{44}})-(\rho _{11}+\rho _{44}).
\end{equation}
This condition can be analyzed for two different physical situations: in the first of them, the ability of achieving a determined entanglement amount for performing the joint-measurement is different in the two laboratories, which means that $\bar{C}_{aA}\neq C_{aA}$.
In the other, the two laboratories are able to achieve the same entanglement amount for carrying out the joint-measurement, that is, $\bar{C}_{aA}=C_{aA}$.

\begin{center}
\emph{Situation B1:} $\bar{C}_{aA}\neq C_{aA}$      
\end{center}

We start by assuming that the channel state $\rho_{AB}$ occupies only the subspace $\mathcal{H}_{0110}$.
In this case, the constraint (\ref{threshX}) is reduced to the simple form
\begin{equation} \label{Eq17}
C_{aA}C_{AB} < \bar{C}_{aA}\bar{C}_{AB}.
\end{equation}
It is interesting to note that this inequality can be satisfied even if $\bar{C}_{AB} < C_{AB}$, of course, with the cost of having $\bar{C}_{aA}>C_{aA}$. This means that then the answer to our question in affirmative; it is centainly possible to achieve $F_X > F_p$ when the noisy channel (\ref{Xstate}) is less entangled than the pure channel (\ref{purechannel}). The requisite to achieve this is that the joint-measurement basis in the second laboratory has an entanglement large enough that $\bar{C}_{aA}> C_{aA}C_{AB}/\bar{C}_{AB} $ is satisfied.
Thus, the entanglement of the measurement basis becomes a key ingredient to attain optimal fidelities when the teleportation is carried out with a mixed channel.
This is the most important result of this work, and it is not restricted to mixed channels $\rho_{AB}\in\mathcal{H}_{0110}$, but it also holds for more general $X$-state channels, as we discuss in the following paragraphs.

Let us go back to the constraint (\ref{threshX}).
This is satisfied when
\begin{equation}
C_{aA}C_{AB} < 1-\mathfrak{C}(1),  \label{Eq26c}
\end{equation}
and one of the two following sets of inequalities are fulfilled,
\begin{subequations}\label{Eq26}
\begin{eqnarray}
\bar{C}_{AB}&>& \frac{C_{aA}}{\bar{C}_{aA}}C_{AB}+\mathfrak{C}(\bar{C}_{aA}),   \label{Eq26a}   \\
\bar{C}_{aA} &>& \frac{\rho _{11}+\rho _{44}+C_{aA}C_{AB}}{1+2\sqrt{\rho _{11}\rho _{44}}},   \label{Eq26b}
\end{eqnarray}
\end{subequations}
or
\begin{subequations}\label{Eq27}
\begin{eqnarray}
\bar{C}_{AB} &>& C_{aA}C_{AB}+\mathfrak{C}(1),   \label{Eq27a}   \\
\bar{C}_{aA} &>& \frac{\rho _{11}+\rho _{44}+C_{aA}C_{AB}}{\bar{C}_{AB}+2\sqrt{\rho _{11}\rho _{44}}}.  \label{Eq27b}
\end{eqnarray}
\end{subequations}
We notice that the threshold values given by the expressions (\ref{Eq26a}) and (\ref{Eq27a}) allow
that $\bar{C}_{AB}$ can be smaller than $C_{AB}$. This is accomplished if
\begin{equation}
C_{AB}C_{aA} < C_{AB}-\mathfrak{C}(1),  \label{Eq28}
\end{equation}
and $\bar{C}_{aA}$ is greater than the greatest threshold, i.e., if (\ref{Eq27b}) is satisfied.
In this way, for a given $\rho_{AB}$, the right hand side of the inequality (\ref{Eq28}) can be interpreted as an upper bound of the entanglement product $C_{AB}C_{aA}$,
in order that the highest fidelity can be attained with the $X$-state channel whenever $\bar{C}_{aA}$ obeys (\ref{Eq27b}).
We notice that the upper bound of the product $C_{AB}C_{aA}$ in Eq. (\ref{Eq26c}) is greater than the one of Eq. (\ref{Eq28}).
Accordingly, in the range $C_{AB}-\mathfrak{C}(1) < C_{AB}C_{aA} < 1-\mathfrak{C}(1)$, the requirement $F_X>F_p$ demands $\bar{C}_{AB} > C_{AB}$.

\begin{figure}[t]
\includegraphics[angle=360, width =0.4\textwidth ]{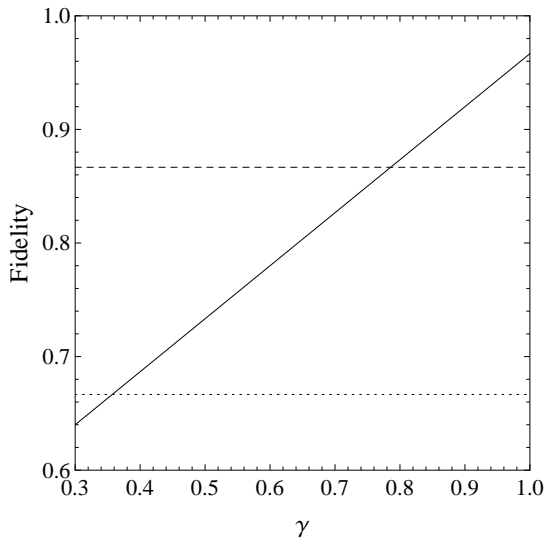}
\caption{The Average fidelity $F_X$ (solid) as a function of $\gamma$.
The references values $F_p=13/15$ (dashed) and the classical limit $2/3$ (dotted).}
\label{Fig1}
\end{figure}

The following example illustrates the arising of the special effect.
The first laboratory performs teleportation with a Bell state channel and the joint-measurement with $C_{aA}=0.6$
In the second laboratory the teleportation is implemented with $\bar{C}_{aA}=0.9$ and a Werner state channel \cite{WS},
\begin{equation}
\rho_{AB}=\frac{1-\gamma}{4}I+\gamma |\tilde{\psi}_{+}\rangle \langle\tilde{\psi}_{+}|,
\end{equation}
which is entangled for $\gamma>1/3$ \cite{APeres}.
In this case $F_p=13/15$ and $F_X=(15+14\gamma)/30$.
The inequalities (\ref{Eq26c}) and (\ref{Eq28}) are satisfied identically, whereas the inequalities (\ref{Eq26a}) and (\ref{Eq27b}) are fulfilled for $\gamma>11/14$.
The Fig. \ref{Fig1} shows the average fidelities, $F_X$ as a function of $\gamma$ and the constant $F_p$.
We can see that $F_X$ becomes greater than $F_p$ just for $\gamma>11/14$.
Besides, $F_X>2/3$ for $\gamma>5/14$.

\begin{center}
\emph{Situation B2:} $\bar{C}_{aA}=C_{aA}$      
\end{center}

We will center in the case $\bar{C}_{aA}\neq C_{AB}$ and $\bar{C}_{aA}\neq \bar{C}_{AB}$.
Under these conditions the constraint (\ref{threshX}) is fulfilled if
\begin{equation}
C_{AB} < 1-\mathfrak{C}(1),   \label{Eq20c}
\end{equation}
and if one of the two following sets of inequalities takes place,
\begin{subequations}\label{Eq20}
\begin{eqnarray}
\bar{C}_{AB} &>& C_{AB}+\mathfrak{C}(\bar{C}_{aA}),   \label{Eq20a}\\
\bar{C}_{aA} &>& \mathcal{C}(1-C_{AB}),   \label{Eq20b}
\end{eqnarray}
\end{subequations}
or
\begin{subequations}\label{Eq21}
\begin{eqnarray}
\bar{C}_{AB} &>& C_{AB}+\mathfrak{C}(1),   \label{Eq21a}\\
\bar{C}_{aA} &>& \mathcal{C}(\bar{C}_{AB}-C_{AB}),   \label{Eq21b}
\end{eqnarray}
\end{subequations}
The inequality (\ref{Eq20c}) defines an upper bound for the concurrence of the pure channel under which we can find an $X$-state which allow to achieve $F_X>F_p$.
We observe that this upper bound vanishes in the special cases for which the subspace $\mathcal{H}_{0011}$ is uniformly or not populated.

The inequalities (\ref{Eq20a}) and (\ref{Eq21a}) state that the concurrence of the noisy channel (\ref{Xstate}) must be higher than both
the entanglement of the pure channel and the threshold values for $F_X$ to show quantum features, as expected; specifically, they are higher than the sum of them.
The inequalities (\ref{Eq20b}) and (\ref{Eq21b}) define threshold values for the concurrence of the measurement basis,
which are always higher than the smallest threshold values $\mathcal{C}(1)$ required for having $F_X>2/3$.
Besides, the greatest of them, $\mathcal{C}(\bar{C}_{AB}-C_{AB})$, is greater than the greatest threshold $\mathcal{C}(\bar{C}_{AB})$.
However, the smallest of them, $\mathcal{C}(1-C_{AB})$, is smaller than the greatest threshold $\mathcal{C}(C_{23})$ if
\begin{equation*}
\bar{C}_{AB} + C_{AB} > 1,
\end{equation*}
otherwise, $\mathcal{C}(1-C_{AB})$ is greater than the greatest threshold $\mathcal{C}(C_{23})$ if
\begin{equation*}
\bar{C}_{AB} + C_{AB} < 1.
\end{equation*}
This shows that sum of the concurrences of the channels $\bar{C}_{AB} + C_{AB}=1$ becomes a critical value over which there is a threshold of the joint measurement,
for improving the average fidelity, smaller than the greater threshold of the joint measured for having $F>2/3$.
In other words, a smaller threshold value of the joint measured demands a greater sum of the entanglement of the channels.

\section{Conclusion}

We have addressed the problem of determining, in the teleportation scheme, the conditions under which a particular entangled noisy channel can be more efficient than a partially entangled pure state channel, even though the former channel is less entangled than latter. To attain this effect it is important to have the capacity of performing a joint measurement with greater entanglement.
We have obtained general formulas for the threshold values and for the upper bounds of the concurrences involved in this approach, finding that $X$-states with the subspace $\mathcal{H}_{0011}$ uniformly or not populated become special channels for which some or all the threshold values and upper bounds vanish.

This is an evidence that the entanglement of the measurement basis plays a role as important as the entanglement of the channel itself, becoming the key factor for optimizing the teleportation fidelity when the only available channel is noisy.
Equivalently, both entanglement works collectively for counteracting the undesired effect produced by the environment.
It is worth to mention that experimentally may be more feasible to manipulate the entanglement of the joint measurement than the channel, with the purpose of improving the teleportation fidelity.

The authors thank grant FONDECyT 1161631.

\end{document}